# Multifunctional Resonant Wavefront-Shaping Meta-Optics Based on Multilayer and Multi-Perturbation Nonlocal Metasurfaces

**Authors:** Stephanie C. Malek[1], Adam C. Overvig[1,2], Andrea Alù[2,3], Nanfang Yu[1,*]

**Affiliations:**
[1]Department of Applied Physics and Applied Mathematics, Columbia University, New York, NY 10027, USA.
[2]Photonics Initiative, Advanced Science Research Center, City University of New York, New York, NY 10031
[3]Physics Program, Graduate Center, City University of New York, New York, NY 10016
[*]Correspondence to: ny2214@columbia.edu

Abstract

Photonic devices rarely provide both elaborate spatial control and sharp spectral control over an incoming wavefront. In optical metasurfaces, for example, the localized modes of individual meta-units govern the wavefront shape over a broad bandwidth, while nonlocal lattice modes extended over many unit cells support high quality-factor resonances. Here, we experimentally demonstrate nonlocal dielectric metasurfaces in the near-infrared that offer both spatial and spectral control of light, realizing metalenses focusing light exclusively over a narrowband resonance while leaving off-resonant frequencies unaffected. Our devices attain this functionality by supporting a quasi-bound state in the continuum encoded with a spatially varying geometric phase. We leverage this capability to experimentally realize a versatile platform for multispectral wavefront shaping where a stack of metasurfaces, each supporting multiple independently controlled quasi-bound states in the continuum, molds the optical wavefront distinctively at multiple wavelengths and yet stay transparent over the rest of the spectrum. Such a platform is scalable to the visible for applications in augmented reality and transparent displays.

Metasurfaces—structured planarized optical devices with a thickness thinner than or comparable to the wavelength of light—typically support a "local" response, i.e., they tailor the optical wavefront through the independent response of each meta-unit. In contrast, "nonlocal" metasurfaces are characterized by an optical response dominated by collective modes over many meta-units (*1, 2*). Local metasurfaces have been widely explored to impart spatially varying phase distributions that shape the impinging optical wavefront to achieve functionalities such as lensing and holography (*3, 4*). However, these devices have typically limited spectral control: since the optical interactions with the meta-units are confined to deeply subwavelength structures, they are typically broadband, and the wavefront deformation is inevitably extended over a wide frequency range (**Fig. 1a** left panel). In contrast, nonlocal metasurfaces, such as guided-mode resonance gratings (*4, 5*) and photonic crystal slabs (PCSs) (*6, 7*), can produce sharp spectral features (**Fig. 1a** middle panel), since they rely on high quality-factor (Q-factor) modes extending transversely over many unit cells. These modes, however, typically cannot at the same time spatially tailor the optical wavefront. Nonlocal metasurfaces hold promise for applications such as sensing (*8, 9*), modulation (*10, 11*) and enhancement of nonlinear optical signals (*12, 13*).

In this work, we design and experimentally realize nonlocal metasurfaces that shape optical wavefronts exclusively at selected wavelengths, leaving the optical wavefront impinging at other wavelengths unchanged (**Fig. 1a** right panel). Our theoretical work has developed the framework of nonlocal metasurfaces that shape the wavefront only on resonance (*14, 15*). This is achievable through a scalable rational design scheme previously only available to local metasurfaces, in which the configuration of scatterers across the surface is determined by reference to a pre-computed library of meta-units. In this rational design scheme, we can devise single-layer nonlocal metasurfaces that shape wavefronts distinctly at different resonances but leave the wavefront shape unchanged at non-resonant wavelengths (**Fig. 1b** right panel). We can also stack multiple nonlocal wavefront-shaping metasurfaces together to attain entirely different functionalities at different wavelengths (**Fig. 1b** left panel). Our theoretical works demonstrated nonlocal metasurfaces with the simple functionalities of single-function cylindrical lensing, multifunctional beam-steering (*14*, *15*) and orbital angular momentum manipulation (*16*, *17*). Here, we design and experimentally realize more complex devices including single-function radial lenses, and multifunctional lenses based on the two approaches illustrated in **Fig. 1b**. Our nonlocal radial metalens explicitly demonstrates two-dimensional (2D) spatial control of the wavefront at the resonant frequency, which is not possible based on recent approaches (*18*) despite the opportunities it presents (*19*).

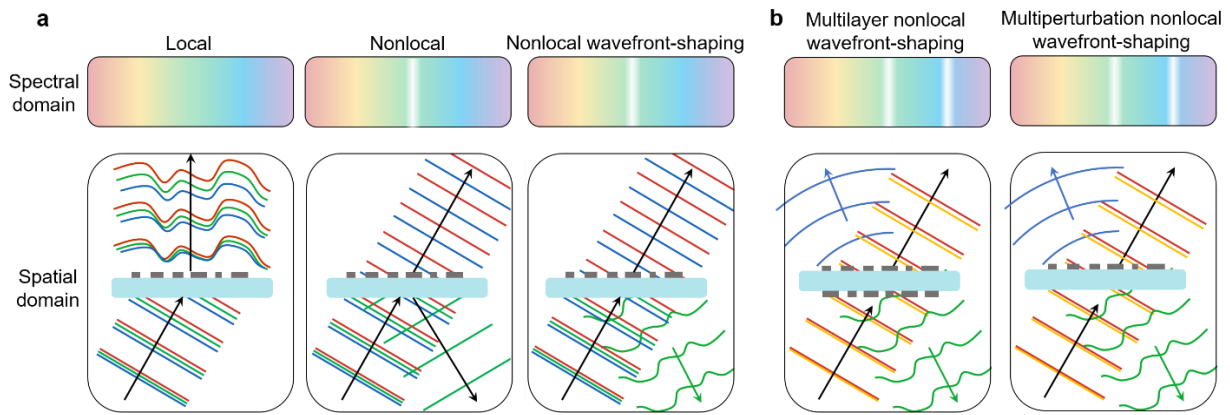

**Figure 1**. Functionality of resonant, wavefront-shaping metasurfaces. (a) Schematic illustrating the distinction between three types of metasurfaces. The nonlocal wavefront-shaping metasurface demonstrated in this work provides spatial control exclusively across its sharp spectral features: it molds optical wavefronts only at the resonant frequency, while leaving the optical wavefronts impinging at other frequencies unchanged. (b) Schematic illustrating two approaches to realize multifunctional nonlocal wavefront-shaping meta-optics.

The operating principles of our nonlocal, wavefront-shaping metasurfaces are rooted in the physics of periodic dielectric PCSs that support leaky extended modes. Such PCSs are known to support bound states in the continuum (BICs), which are modes with infinite radiative quality-factors (Q-factors) despite being momentum-matched to free space (*8*). Applying a perturbation to break in-plane inversion symmetry of meta-units may create a quasi-BIC (q-BIC) that is leaky and excitable from free space light (*21*). Alternatively, a leaky state may be formed using a dimerizing perturbation (i.e., a perturbation that doubles the period along a real-space dimension and halves the first Brillouin zone) that folds a previously guided mode into the radiation continuum, also realizing a q-BIC. In either case, both the scalar and vectorial properties of q-BICs can be readily engineered, and can be controlled with extreme precision through the perturbation. The scalar property of optical lifetime or Q-factor is controlled by the magnitude of the perturbation $\delta$ as $Q \propto \frac{1}{\delta^2}$ (*20*, *21*). The vectorial property of polarization is controlled by the type of perturbation, which we have catalogued in our previous theoretical work (*14*). A key implication of the vectorial properties of q-BICs is that specific symmetry groups impart a geometric phase to circularly polarized light because the linear polarization of the q-BIC varies directly with the in-plane rotation angle of the perturbation $\alpha$ (*15*). Specifically, in *p2* plane groups—a lattice with two-fold symmetry—the geometric phase of transmitted light with converted handedness of circular polarization experiences a geometric phase of $\Phi \sim 4\alpha$ exclusively at narrowband q-BIC resonances. Crucially, the linear polarization of the q-BIC radiation is locally orientable with respect to the fixed global lattice by changing $\alpha$ (**Supplementary Section 2**), enabling the 2D spatial phase profiles of the radiative component of the q-BIC to be encoded by the perturbation while minimally affecting the nonradiative properties of the q-BIC (e.g., near-field mode profile and resonant frequency). In contrast, recent work on resonant phase gradient meta-gratings with high-Q (*18*) have been limited to high deflection angles in a single fixed direction relative to the grating. While geometric phase in q-BICs is fundamentally robust because it derives from symmetry, the meta-unit library must be designed carefully to produce devices that exert accurate spectral and spatial control over the wavefront when fabricated with typical nanofabrication imperfections. **Figure S1** details an experimentally workable *p2* q-BIC geometric-phase meta-unit library in the near-infrared on a substrate of amorphous silicon on glass, and **Fig. S2** explores the associated design considerations of the unperturbed PCS and its perturbation.

We experimentally demonstrate (**Figs. 2a-c**) a nonlocal radial metalens with NA=0.2 and a diameter of 800 μm using the meta-unit library in **Figure S1**. This metalens has a resonance centered at $\lambda$=1,590 nm with a Q-factor of ~86. The wavelength exclusivity of this feature follows closely with the theoretical expectation: A series of transverse two-dimensional (2D) far-field scans shows that focusing is most efficient at the center of the resonance, $\lambda$=1,590 nm, with the focusing efficiency dropping at the two shoulders of the resonance, $\lambda$=1,575 nm and 1,600 nm, and that the focal spots become almost undetectable at wavelengths tens of nanometers away from the center of the resonance (**Fig. 2e**). Longitudinal 2D far-field scans of the device (**Fig. 2f**) reveal that the focal spots at resonance ($\lambda$=1,575-1,600 nm) are orders of magnitude brighter than the focal spots off resonance, following a Lorentzian line shape. The device is functionally transparent off resonance in that the background planewave is estimated to be three to four orders of magnitude stronger in power than the focal spots at off-resonance wavelengths. Notably, the focal spot at resonance is diffraction limited: vertical and horizontal linecuts of the focal spot at the center of the resonance (**Fig. 2g**) reveal Strehl ratios (estimated from the Airy disc and first ring of the intensity pattern) of 0.89 and 0.85 in the x and y directions, respectively. That the lens is diffraction limited suggests that the resonant wavelength is nearly constant across the entire device, else the effective numerical aperture at the resonant wavelength would be smaller than predicted. At the peak of the resonance the device exhibits a maximum conversion efficiency of ~8% of the incident power, as indicated by the transmission spectra in **Fig. 2d**. Note that this conversion efficiency is

~32% of the theoretical maximum (which is 25% of the incident power) (*15, 22*) and that chiral nonlocal metasurfaces are needed to achieve 100% efficiency (*23*).

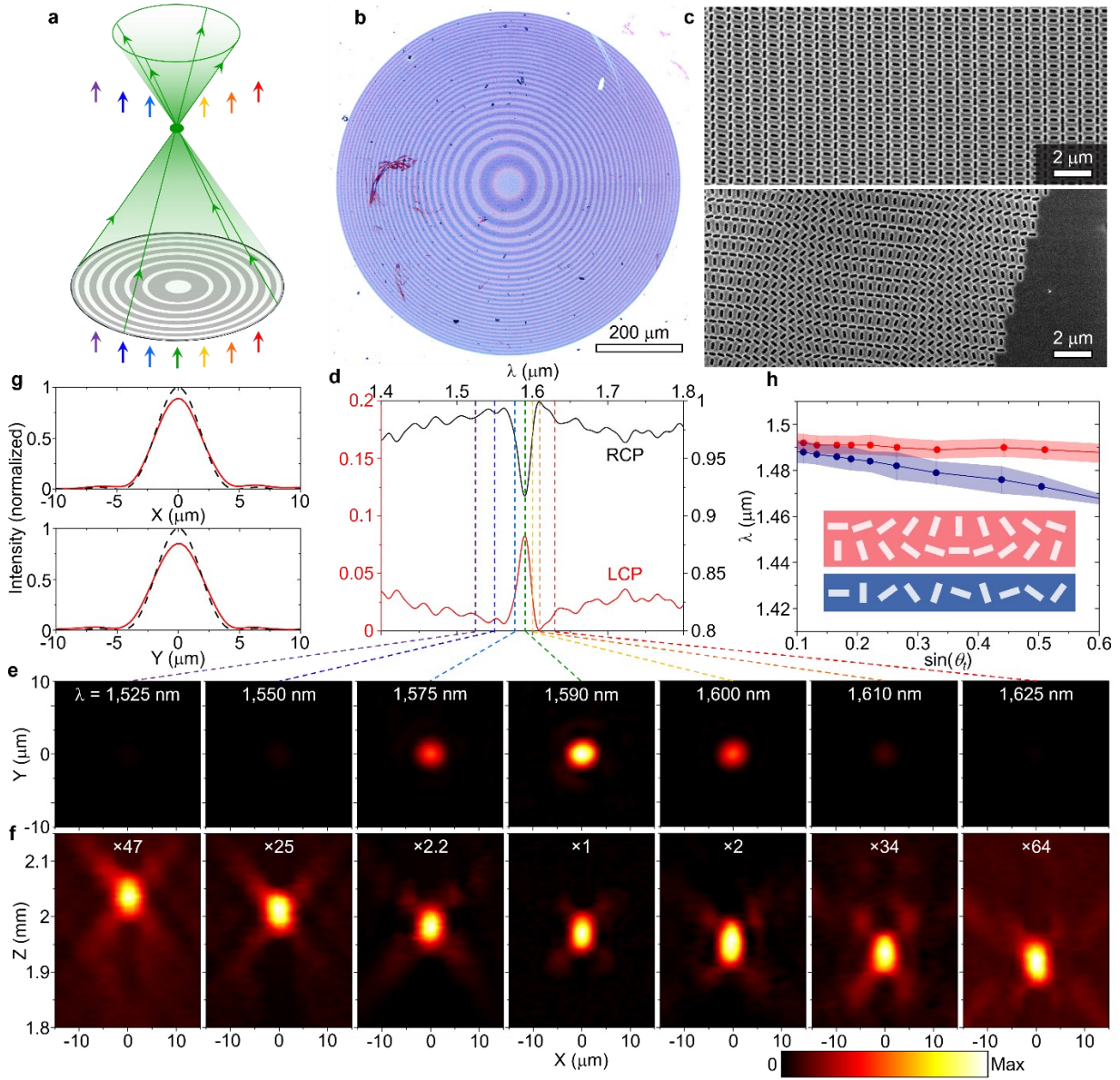

**Figure 2.** Experimental results of a resonant radial metalens with NA=0.2. (a) Illustration showing the resonant operation of the metalens, with 'red' light being focused, while the other colors are passed without distortion. (b) Photograph of the metalens with a diameter of 800 μm. (c) Scanning electron microscope (SEM) images of the device at its center (top) and edge (bottom). (d) Measured transmission spectra of the metalens for light of converted and unconverted handedness of circular polarization. (e) Measured transverse intensity distributions on the focal plane. (f) Measured longitudinal intensity distributions on a plane through the focal spot. The metalens is located at Z=0. (g) Measured (solid red curves) and theoretical (black dashed curves) linecuts of the focal spot at the center of the resonance, λ=1,590 nm, along the x and y directions. (h) Simulated resonant wavelength dispersion as a function of refraction angle for phase-gradient nonlocal metasurfaces excited by normally incident light. Dots represent the center of a resonance and shaded regions FWHM of the resonance. Data for the cases in which the phase gradients are orthogonal to the direction of dimerization perturbation are shown in red and those for the cases in which the phase gradients are along the direction of the perturbation are shown in blue. Inset: Schematics illustrating the two alignments between the phase gradient and the dimerization perturbation.

Keeping the entire metalens resonant is nontrivial. The unperturbed lattice of the nonlocal metasurface is described by a band structure; consequently, the resonant wavelengths of the nonlocal modes are dispersive with the deflection angle or phase gradient implemented by the metasurface (**Fig. 2h**) (*14*, *15*). This represents a key design constraint: the total shift in resonant wavelength due to the phase-gradient variation across a device must be smaller than the full width at half maximum (FWHM) of the resonance. As such, given a certain band curvature, there is a tradeoff between the Q-factor and the range of deflection angles supportable across a device, commonly manifested as the numerical aperture (NA) of the metalens (*14*, *15*). The curvature of the most dispersive direction limits the maximum achievable NA. For the simulated meta-unit library of the radial metalens, the estimated maximum NA for a radial metalens is ~0.26 (**Fig. 2h**). One pathway towards realizing high NA devices is composing a radial lens with slices of cylindrical lenses using only the least dispersive relative orientation between the phase gradient and the dimerization perturbation (**Supplementary Section 3**). The ultimate solution is through bandstructure engineering of PCSs (*24*, *25*): a flatter band will allow for the creation of devices with simultaneously higher Q-factors and larger NA.

Further advances with nonlocal metasurfaces can be achieved by developing multispectral nonlocal meta-optics—either by cascading nonlocal metasurfaces with distinct resonant frequencies, or by adding to a single-layer metasurface a set of orthogonal perturbations, each of which imparts an independent geometric phase profile (*15*). We pursue both approaches here, and then show compound meta-optics combining both methods. Beginning with the first approach, we experimentally demonstrate a nonlocal metalens doublet that focuses light at two selected wavelengths and leaves the wavefront shape unchanged at non-resonant wavelengths. This doublet consists of a converging cylindrical lens with NA=0.1 resonant at a shorter wavelength of $\lambda$=1,450 nm and a diverging radial lens with NA=0.2 resonant at a longer wavelength of $\lambda$=1,590 nm. They are arranged so that they share the same focal plane located between the two elements (**Fig. 3a**) but may be rearranged as desired. Both elements are devised from meta-unit libraries of rectangular apertures etched in a 125-nm thick silicon film on glass for convenience, but each element could be based on a distinct material platform or with a different meta-unit motif for more advanced functionalities. The radial lens is the same device as **Fig. 2**, acting here as a diverging lens because the handedness of circularly polarized incident light has been reversed. Compared to this design, the meta-units for the cylindrical lens, as detailed in **Fig. S3**, have smaller dimensions to blueshift the resonant wavelength to $\lambda$=1,450 nm (**Fig. 3b**). Multiwavelength transverse far-field scans at the focal plane (**Fig. 3c**) show that at $\lambda$=1,450 nm, one element of the doublet (the cylindrical lens) generates a focal line, while at $\lambda$=1,600 nm, the other element of the doublet (the radial lens) produces a focal spot. Off resonance, there is minimal transmission of handedness-converted light: a plane wave transmits through the doublet with no polarization conversion nor wavefront deformation. Due to the broadband transparency, this is a flexible platform where many device configurations can be realized. For instance, **Fig. S11** shows a metasurface doublet where the metalenses do not share a focal plane. Additional functionalities may also be attained by cascading more than two metasurfaces.

In contrast, it is highly nontrivial to cascade conventional local metasurfaces to achieve multifunctionality. To readily cascade metasurfaces and attain different functionalities at different wavelengths, individual metasurfaces must not shape the wavefront at their non-designed wavelength (**Figure S8**). In our approach, it is trivial to cascade our nonlocal metasurfaces if their q-BICs do not spectrally overlap. The resulting stack performs the collective functions of all the individual metasurfaces and does not alter the wavefront shape at non-resonant wavelengths. However, local metasurfaces rarely satisfy this condition, so simply stacking them can result in distorted wavefronts and degraded functionalities (**Figures S9, S10**). The stack—rather than individual metasurfaces—must be designed as one entity to perform the desired set of functions (**Table S2**). Such systems often prove computationally intensive to devise (*26–28*), require precise lateral alignment of the constituent metasurfaces (*27, 29*), and do not support closely spaced operating wavelengths (*29, 30*).

We also experimentally demonstrate a two-function metasurface based on orthogonal perturbations where each perturbation independently controls one targeted q-BIC. The designed meta-units (**Fig. 3k**) are formed by two sets of rectangular apertures in a silicon thin film, and each set of apertures controls the phase of one q-BIC but not that of the other. In isolation, each set of apertures belong to the *p2* plane group, but together the composite meta-units in this device have only translational symmetry and therefore belong to the *p1* plane group (*15*). This follows the principle of successive orthogonal perturbations (*15*), which is fundamentally distinct from that of spatial multiplexing commonly employed in local (*31, 32*) and nonlocal metasurfaces (*33, 34*) for multifunctionality. Simply multiplexing *p2* meta-units with different operating wavelengths does not achieve the desired multifunctionality (**Section S8**). We note that this meta-unit library (**Fig. 3l-m**) is designed to have lower Q-factors than the multi-perturbation device in our previous theoretical work (*15*) in order to implement nonconstant phase gradients and to improve the device robustness against fabrication imperfections. Specifically, the Q-factor varies inversely with the magnitude of the perturbation (*20*), but simply increasing the latter (e.g., more highly anisotropic apertures) is not always a sufficient or experimentally practical method to achieve a target Q-factor. So in contrast to the purely symmetry driven designs in our initial theoretical proof of principle demonstration (*15*), here the motif and height of the meta-units are judiciously chosen.

We devise a proof-of-concept metasurface implementing two orthogonal cylindrical lenses, both with NA=0.05, where the phase profile for each lens is controlled by only one set of perturbations (**Fig. 3g**). An optical microscope image of the fabricated device (**Fig. 3h**) shows the horizontal and vertical zones of the two lenses overlapping within the single metasurface as a result of the two independently tiled sets of apertures (SEM images: **Fig. 3i-j**). Imaging handedness-converted light from this device at the shared focal plane reveals a horizontal focal line at $\lambda$=1,385 nm, a vertical focal line at $\lambda$=1,460 nm, and mostly flat wavefronts at non-resonant wavelengths (**Fig. 3n**). We have therefore demonstrated that both approaches to multifunctional nonlocal meta-optics—cascaded metasurfaces and multiple independent perturbations—are achievable experimentally, with cascading leaving open the possibility of reconfiguring the constituent metasurface elements and multiple perturbations allowing for a thin single-layer, multifunctional device. In principal, up to four distinct functions (e.g., phase profiles) from four distinct q-BICs may be realized on a single metasurface (*15*), at the cost of denser patterning and increased cross-talk.

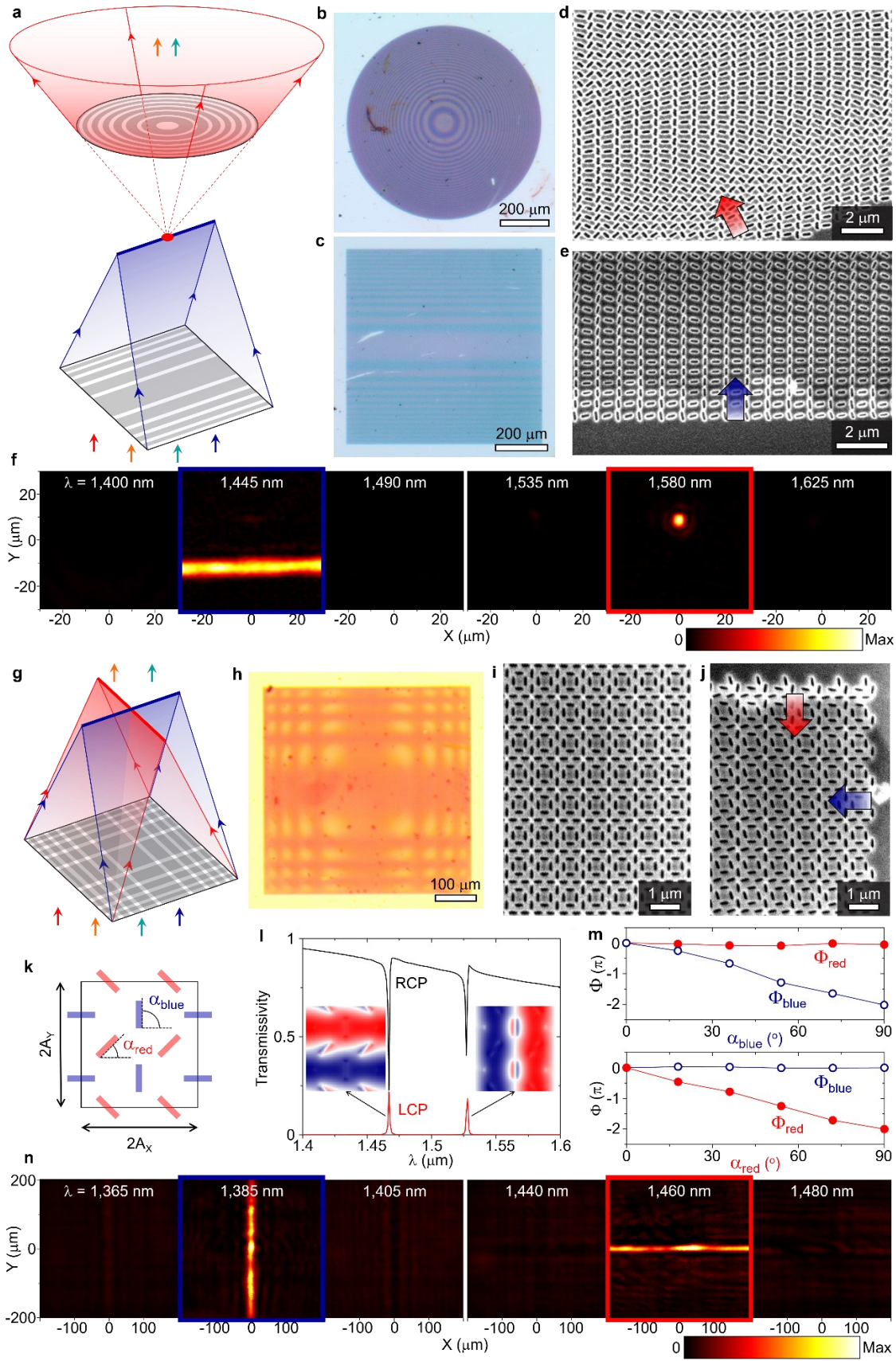

**Figure 3.** Experimental results of multifunctional meta-optics devised by a multilayer approach (a-f) and a multi-perturbation approach (g-n). (a) Schematic illustrating the multilayer approach with a cylindrical lens with NA=0.1 made from the meta-unit library in **Fig. S3** as the converging element and the radial lens from **Fig. 2** as the diverging element. (b-c) Photographs of the diverging (b) and converging (c) elements of the doublet. (d-e) SEM images of the diverging (d) and converging (e) elements. Arrows indicate phase gradient direction. (f) Measured transverse intensity distributions of transmitted light of converted handedness of the doublet. (g) Schematic illustrating the multi-perturbation approach for a dual-function cylindrical lens. (h) Optical image of the device. (i-j) SEM images of the device at the center (i) and corner (j) of the device. (k) Schematic of a meta-unit of the dual-function cylindrical lens. Shown are two sets of apertures (colored red and blue) defined in a silicon thin film with a thickness of H=200 nm; lattice constants are $A_x$=430 nm $A_y$=460 nm, and aperture dimensions are W=50 nm, L= 200 nm. (l) Simulated transmission spectra of light with converted (red) and preserved (black) handedness. Insets: Out of plane component of electric field for the two modes. (m) Geometric phases of both modes as a function of in-plane rotation angle of the blue (top) and red (bottom) perturbations, respectively. (n) Measured transverse intensity distributions of handedness converted light on the focal plane of the dual-function cylindrical lens. Device dimensions are detailed in **Table S1**.

These two approaches to multispectral nonlocal meta-optics can be combined to experimentally realize highly multifunctional meta-optics. We begin by assembling a three-function doublet (**Fig. 4a**) by stacking a dual-function cylindrical lens as a converging element (**Fig. 4 b, d, e**) and a single-function quasi-radial lens as a diverging element (**Fig. 4 c, f, g**) such that the two metasurfaces share a focal plane. The measured transverse intensity distributions on the shared focal plane reveal a horizontal focal line at λ=1,428 nm and a vertical focal line at λ=1,486 nm from the dual-function cylindrical lens, a focal spot from the quasi-radial lens at λ=1,620 nm, and minimal handedness-converted light at other wavelengths (**Fig. 4h**). The Q-factors for these three lensing functions range from ~60 to ~90 (**Fig. 4i**).

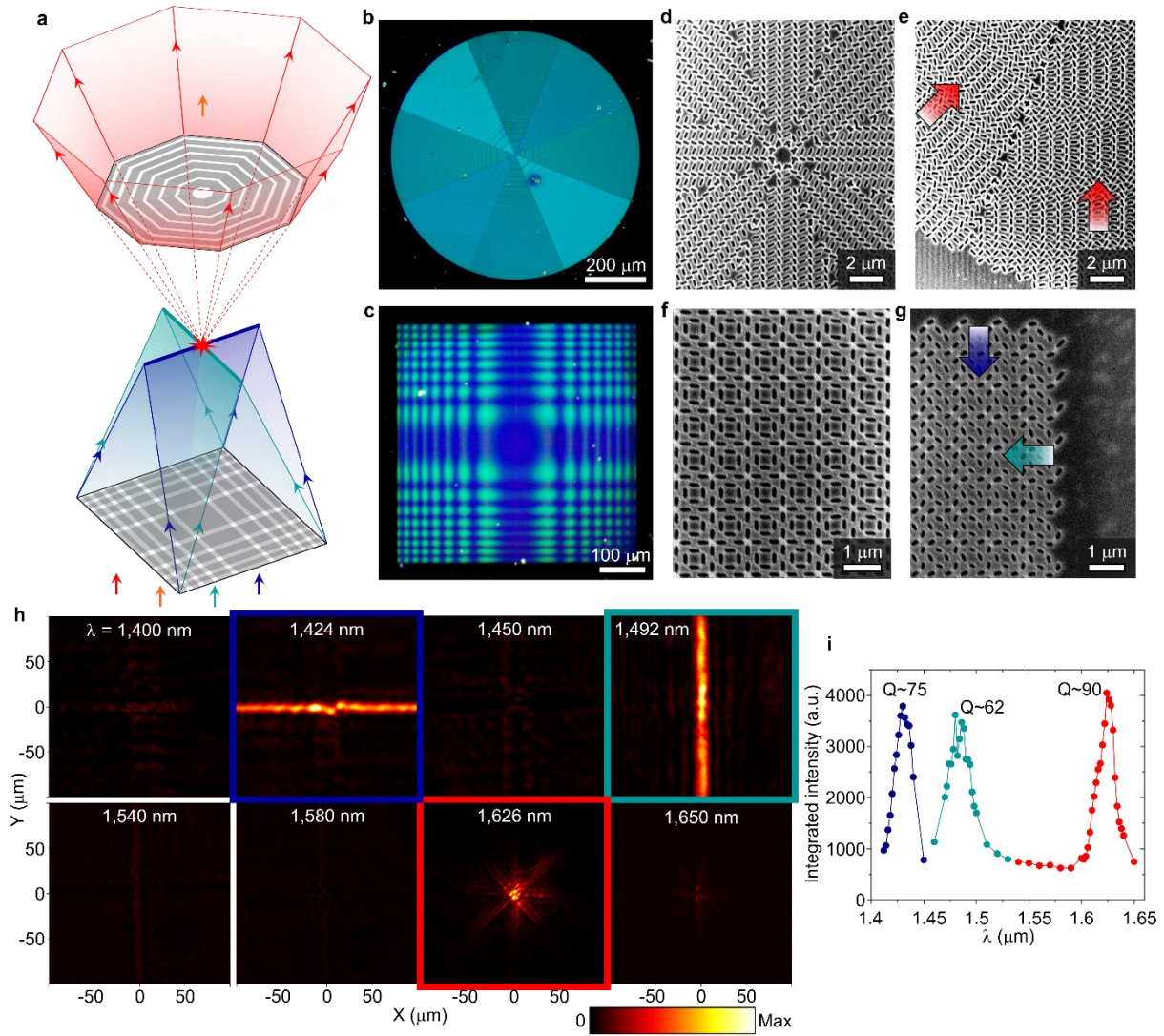

**Figure 4.** Experimental results of a three-function doublet. Device dimensions are detailed in Table S1. (a) Schematic illustrating the operation of the three-function doublet with a single-function quasi-radial lens with NA=0.4 made from the meta-unit library in **Fig. S1** as a diverging element and a dual-function metalens (two orthogonal cylindrical lenses with NA~0.1 and dimensions H=100 nm, $A_x$=430 nm, $A_y$=460 nm) as a converging element. (b-c) Dark field optical microscope images of the diverging (b) and converging (c) devices. (d-e) SEM images of the diverging device at its center (d) and edge (e). (f-g) SEM images of the converging device at its center (f) and edge (g). Arrows indicate phase gradient direction. (h) Measured transverse intensity distributions of handedness-converted light of the doublet at the shared focal plane. (j) Estimated Q-factors for the three lensing functions by integrating the optical intensity near the focal features (focal lines or focal spot) as a function of wavelength (**Supplementary Section 8**).

Further functionality can be realized by cascading more than one multifunctional metasurface, such as a four-function doublet (**Fig. 5a**) consisting of two dual-function cylindrical lenses (**Fig. 5b-e**) that share the same focal plane. The measured transverse intensity distributions on the shared focal plane show four distinct focal lines—at λ=1,414 nm and 1,622 nm from the converging element and λ=1,388 nm and 1,460 nm from the diverging element (**Fig. 5f**). The Q-factors for these four lensing functions range from ~100 to ~300 (**Fig. 5g**). Our demonstration of highly multifunctional meta-optics features more distinct functionalities per wavelength range and per number of metasurface layers than has been previously reported for cascaded local metasurfaces (**Table S2**). The degree of multifunctionality can be largely boosted by stacking many nonlocal metasurfaces each with up to four independent perturbations (*15*).

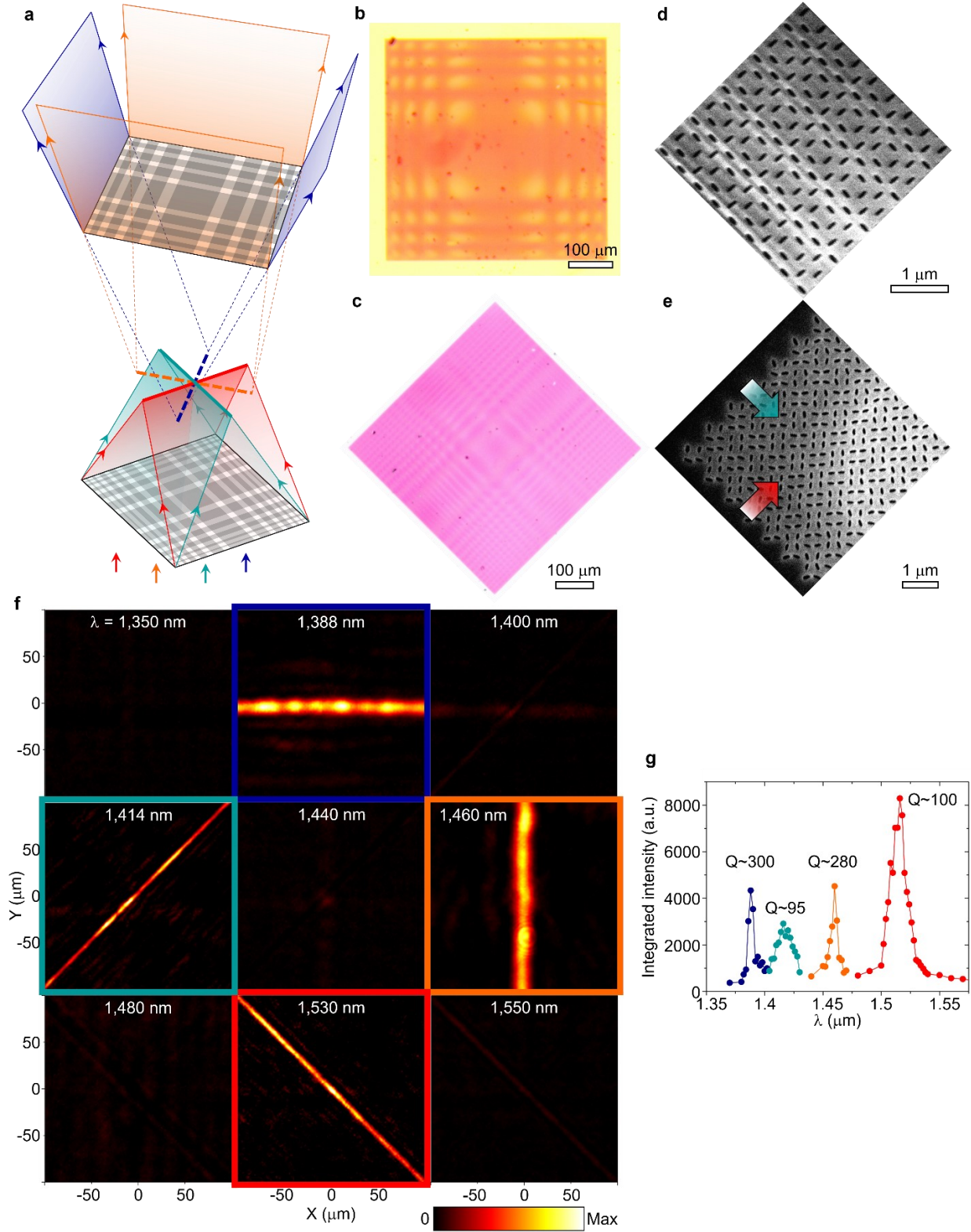

**Figure 5**. Experimental results of a four-function doublet. Device dimensions are detailed in Table S1. (a) Schematic of the doublet comprised a two-function set of cylindrical lenses with NA~0.14

and H=100 nm as the converging element and the multi-perturbation device in **Fig. 3** with NA~0.05 as the diverging element. (b-c) Bright field optical microscope images of the diverging (b) and converging (c) elements. (d-e) SEM images of the diverging element at its center (d) and converging element at its corner (e). Arrows indicate the direction of the phase gradient. (f) Measured transverse intensity distributions for the four-function doublet at the shared focal plane. (g) Estimated Q-factors for the three lensing functions by integrating the optical intensity near the focal features (focal lines or focal spot) as a function of wavelength (**Supplementary Section 8**).

The experimental demonstration of these multifunctional devices is challenging compared to local metasurfaces as a near-constant resonant wavelength must be maintained across the entire device. In addition to the aforementioned angular dispersion, the resonant frequency could also be affected by the quality of device fabrication. For example, the spatial profiles of the modes in the diverging element in **Fig. 5** are such that in-plane components of the electric field are dominant and concentrated within the apertures (**Fig. S14**). As a result, the resonant wavelengths can vary due to small systematic variations in aperture size across the footprint of the device caused by systematic variations in thin-film deposition, lithographic exposure, or etching conditions. In fact, in devices with large fabrication errors, the q-BICs cannot be excited across their footprint by a monochromatic excitation (e.g., devices regionally resonate with incident light based on the frequency) (**Fig. S15**). Finally, we demonstrate in simulations the utility of our multifunctional nonlocal metasurfaces in augmented reality (AR) glasses operating in the visible. Due to their small form-factor and potentially expanded functionality compared to conventional optical components, metasurfaces have attracted growing interest for applications in AR headsets (*35–38*). Thus far, most reported metasurfaces for AR require extra optical components, such as polarizers or beam splitters, that inevitably attenuate real-world light and add size and mass to the headset (**Table S3**). **Figures 6a** and **6b** schematically show our nonlocal metasurface acting as an optical see-through lens that reflects virtual information to the viewer's eye at selected narrowband wavelengths but permits an unobstructed broadband view of the real world. This paradigm allows for a wide field-of-view of virtual information with the nonlocal metasurface covering the entire eyeglass, and does not require extra polarizers or beam-splitters that attenuate real-world light.

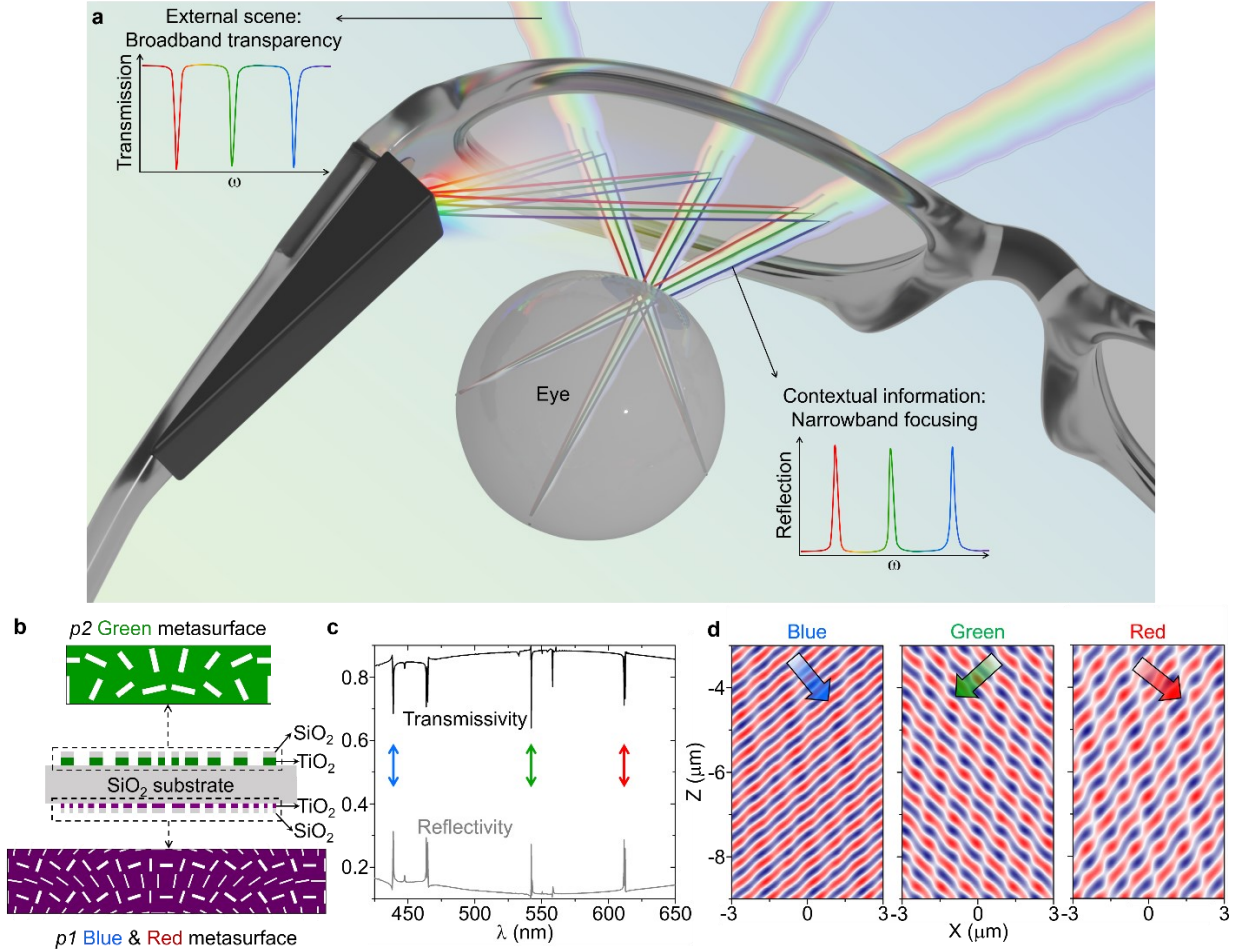

**Figure 6**. Conceptual demonstration of augmented reality enabled by multifunctional nonlocal metasurfaces. (a) Illustration showing the operation of an augmented reality headset with multifunctional nonlocal metasurfaces as optical see-through lenses. (b) Schematic of a super-period of a nonlocal meta-optics system implementing three distinct phase gradients at three chosen visible wavelengths, respectively. See **Table S1** for detailed design parameters. (c) Simulated transmission and reflection spectra of the meta-optics system. Arrows indicate the three colors. (e) Simulated reflected wavefronts shaped by independent q-BICs at $\lambda_B$=438 nm, $\lambda_G$=542 nm, and $\lambda_R$=611.5 nm. Arrows indicate light propagation directions at the three colors.

As a proof-of-concept, we design and numerically demonstrate a nonlocal meta-optics system on a single glass substrate that independently controls the anomalous reflection of three colors of virtual information while remaining transparent to impinging light from the real world (**Fig. 6a**). This design entails a doublet with a single-function metasurface based on *p2* meta-units operating at the green wavelength and a dual-function metasurface based on *p1* meta-units operating at the red and blue wavelengths (**Fig. 6b**). Both metasurfaces are composed of rectangular apertures etched into a thin film of $TiO_2$ covered with an antireflection layer of $SiO_2$ and are compatible with previously demonstrated fabrication methods (*39*). **Figure 6c** highlights the broadband high-transmission of real-world light and the narrowband reflection at the three chosen wavelengths in the visible for the doublet, calculated by finite difference time domain (FDTD) simulations and an incoherent transfer matrix method (*40*). The simulated wavefronts of reflected light on resonance are independently and deliberately steered by the doublet (**Fig. 6d**). We envision future work experimentally realizing nonlocal meta-optics in the visible that also support flat bands to provide high NA to enable practical AR solutions.

In summary, we have experimentally demonstrated nonlocal wavefront-shaping metasurfaces and meta-optics systems including a nonlocal radial metalens, a dual-function cylindrical metalens, and metalens doublets with up to four distinct functionalities. This platform of nonlocal metasurfaces readily allows for independent control of resonant wavelengths (via meta-unit geometry), Q factors (via perturbation strength), resonant frequency dispersion (via bandstructure engineering), and wavefront (via spatial distribution of geometric phase) at a plurality of wavelengths (via cascading and/or adding independent perturbations). These devices may expand the capabilities of multifunctional meta-optics to include active or nonlinear wavefront shaping by leveraging the enhanced light-matter interactions of the high Q-factor, wavefront-shaping resonances (*41*). In addition, incorporating local metasurface design considerations may allow us to realize multifunctional wavefront generation from structured thin films driven by incoherent emitters (*17*). Scaled to visible wavelengths, our resonant metasurfaces may prove useful for AR and transparent display applications as compact multi-color see-through optics.

**Methods**

**Device fabrication** Approximately 100, 125, or 200nm of amorphous silicon (a-Si) is deposited on a fused silica wafer by plasma enhanced chemical vapor deposition (Oxford Instruments NPG90 PECVD). The wafer is spin-coated with poly(methyl methacrylate) (PMMA A4 950) for 45 seconds at 2000 RPM and then baked at 180 °C for 2 minutes. Then an anti-charging layer, DisCharge H2O (DisChem, Inc.), is spun at 2000 RPM for 45 seconds. Devices are patterned with electron beam lithography (Elionix ELS-G100) at a current of 1 nA for the radial lenses and 2 nA for the cylindrical lenses after appropriate proximity effect correction is applied (BEAMER). After exposure, the anti-charging layer is removed by rinsing in DI water, and the devices are developed in a chilled 3:1 isopropyl alcohol:deionized water solution for 2 minutes followed by 30 seconds of rinsing in deionized water. The devices are etched in a fluorine-based inductively coupled plasma etcher (Oxford Instruments PlasmaPro 100). The PMMA etch mask is stripped by soaking the wafer in N-Methyl-2-pyrrolidone (NMP) at 80 °C for several hours.

**Transmission measurements** Transmission spectra are measured with a Fourier transform infrared (FTIR) spectrometer (Bruker Vertex 70v) and a mid-infrared microscope (Bruker Hyperion 2000) with two circular polarizers (Thorlabs) in the beam path—the first one circularly polarizes the incident light and the second selects the handedness of the light transmitted through the device. Raw data is normalized with the following scheme to present polarization converted ($T_c$) and unconverted ($T_u$) transmission spectra.

$$A = converted\ polarization\ on\ device - converted\ polarization\ on\ unpatterned\ wafer$$
$$B = unconverted\ polarization\ on\ device + A$$
$$T_c = \frac{A}{B}$$
$$T_u = \frac{unconverted\ polarization\ on\ device}{B}$$

The above procedure will yield more accurate spectra as the infrared objectives used in our spectroscopic setup introduce some degree of background polarization conversion. Normalized transmission spectra are filtered with a fast Fourier transform filter to remove fine fringes due to Fabry–Pérot interference of light in the substrate.

**Far Field Intensity Scans** The focusing performance of the metalenses is measured according to the framework described in our previous work (*42*). Near-infrared light is coupled from a super continuum source (NKT Photonics SuperK Extreme) through a monochromator (Horiba iHR550) to a fiber collimator and then circularly polarized by properly orienting a linear polarizer and a quarter-wave plate. Light transmitted through the metasurface is collected by a 10× objective and one of its circularly polarized components is selected by another quarter-wave plate and linear polarizer before it is imaged by a near-infrared camera (NIRvana InGaAs camera, Princeton Instruments). The objective, analyzing polarizer, quarter-wave plate, and camera are all mounted on a motorized linear translation stage. With the latter, two-dimensional transmitted intensity patterns are imaged over a 1-2 mm longitudinal range around the focal plane of the metalenses in steps of 1-5 μm at a number of selected wavelengths. A three-dimensional intensity pattern can be created by stacking a series of such two-dimensional intensity patterns.

**Acknowledgements**

This work was supported by the National Science Foundation (grant no. QII-TAQS-1936359 and no. ECCS-2004685) and the Air Force Office of Scientific Research (grant no. FA9550-14-1-0389 and no. FA9550-16-1-0322). S.C.M. acknowledges support from the NSF Graduate Research Fellowship Program (grant no. DGE-1644869). A.C.O. acknowledges support from the NSF IGERT program (grant no. DGE-1069240). Device fabrication was carried out at the Columbia Nano Initiative cleanroom, and at the Advanced Science Research Center NanoFabrication Facility at the Graduate Center of the City University of New York.

**Author Contributions Statement**

S.C.M. and N.Y. conceived the experiments. S.C.M. and A.C.O. conducted analytical calculations and full-wave simulations to design the devices. S.C.M. fabricated the devices, constructed the experimental setup, and characterized device performance. S.C.M., A.C.O. and N.Y. analyzed the data. A.A. and N.Y. supervised the project. All authors prepared and edited the manuscript.

**Competing Interests**

The authors declare no competing interests.